\documentclass{article}
\usepackage{ijcai26}
\usepackage{adjustbox}
\usepackage{times}
\usepackage{soul}
\usepackage{url}
\usepackage[hidelinks]{hyperref}
\usepackage[utf8]{inputenc}
\usepackage[small]{caption}
\usepackage{graphicx}
\usepackage{amsmath}
\usepackage{amsthm}
\usepackage{booktabs}
\usepackage{algorithm}
\usepackage{algorithmic}
\usepackage[switch]{lineno}
\usepackage{makecell}
\usepackage{bigstrut,bigdelim,multirow}

\title{GMENet: Generative Mixture of Experts Network for Multi-Center Glioma Diagnosis with Incomplete Imaging Sequences}

\author{
	Paper ID: \#89
}

\author{
	Pengfei Song$^1$\and
	Fangjin Liu$^{2}$\and
	Wenwen Zeng$^1$\and
	Yonghuang Wu$^1$\and
	Chengqian Zhao$^1$\and
	Feiyu Yin$^{1}$ \and
	Xuan Xie$^{1}$ \And
	Jinhua Yu$^{1,3}$ \thanks{Corresponding Author}
	 \\
	\affiliations
	$^1$School of Biomedical Engineering and Technology Innovation, Fudan University\\
	$^2$Institute of Science and Technology for Brain-Inspired Intelligence, Fudan University\\
	$^3$Intelligent Diagnosis and Treatment Laboratory for Brain Diseases, Joint Laboratory of Neurosurgery Department of Huashan Hospital and School of Information Science and Technology, Fudan University\\
	\emails
	\{23110720139, fjliu24, 22110720094, 24110720122, cqzhao21, 20110720072\}@m.fudan.edu.cn,
	\{20110720072, jhyu\}@fudan.edu.cn
	}

\begin{document}

\maketitle

\begin{abstract}
Contemporary glioma diagnosis integrates molecular features with histopathology to guide clinical decision-making. However, in clinical settings, divergent imaging protocols result in incomplete MRI sequences, leading to two primary challenges: forcing existing frameworks to discard a large portion of clinical data during training and consequently limiting their clinical applicability. To address these limitations, we propose GMENet, a Generative Mixture of Experts Network for multi-center glioma diagnosis with incomplete imaging sequences. Firstly, we design a Cross-attention-based Gated Generation Module that synthesizes missing sequence features from available sequences via cross-attention and dynamic gating mechanisms, incorporating a cycle-consistency loss to preserve semantic integrity. Secondly, we introduce a Dynamically Weighted Experts Fusion Module that performs mixture-of-experts interaction and confidence-aware fusion over original and synthesized dual-sequence features for multi-task prediction. We evaluate GMENet on a multi-center cohort of 1,241 subjects from four in-house datasets and two public repositories. Experiments show that GMENet expands clinically usable training data by 97\%, relative to complete-sequence-only data. Furthermore, it consistently outperforms state-of-the-art methods trained on complete data, demonstrating improved robustness under cross-center distribution shifts. Code is available at: \url{https://github.com/spf-sd/GMENet}.
\end{abstract}

\section{Introduction}
 
\footnote{Generative AI was utilized solely for linguistic polishing. The authors have verified the accuracy of the content and assume full responsibility.} Glioma is the most common malignant tumors of the Central Nervous System (CNS), accounting for approximately 80$\%$ of all CNS malignancies \cite{messali2014review}. Among them, diffuse gliomas represent the majority of adult primary brain tumors. The fifth edition (2021) of the World Health Organization Classification of Tumors of the Central Nervous System (WHO CNS5) stratifies diffuse gliomas into three categories: (1) oligodendroglioma, IDH-mutant and 1p/19q co-deleted; (2) astrocytoma, IDH-mutant; and (3) glioblastoma, IDH wildtype \cite{louis20212021}. Importantly, WHO CNS5 formalizes an integrated diagnostic paradigm by combining specific molecular features (IDH mutation status and 1p/19q co-deletion) with histopathological assessment. This paradigm highlights the critical role of molecular features in tumor biology, therapeutic decision-making, and prognosis \cite{eckel2015glioma,nobusawa2009idh1}. This shift marks the era of integrated diagnosis in the clinical evaluation of gliomas.

MRI is the most widely used imaging modality for glioma assessment \cite{cheng2021multimodal}, offering high-resolution, noninvasive visualization with substantial diagnostic value. Among the sequences commonly employed for preoperative evaluation, T1-weighted contrast-enhanced (T1c) imaging visualizes the tumor core, while T2-fluid-attenuated inversion recovery (FLAIR) sequences effectively delineate the tumor infiltration zone. 

Most current methods \cite{cheng2022fully,choi2021fully,wu2022swin,van2023combined,zhang2024biologically} have shown promise for glioma molecular subtyping and pathology-type prediction, but they are designed and optimized under the assumption of complete sequences. Contrary to this ideal assumption, in real-world multi-center clinical practice, data from different institutions often exhibit incomplete MRI sequences due to factors such as differences in imaging protocols, inconsistent scanning procedures, or equipment limitations. Incomplete sequences force many frameworks to discard incomplete samples to fit the model, resulting in the waste of a large amount of valuable clinical data and limiting the robustness and clinical applicability of the model under cross-center distribution shifts.

To address these challenges and explore potential solutions, we present a novel GMENet (Generative Mixture of Experts Network) for multi-center glioma diagnosis with incomplete imaging sequences. The core objective of GMENet is to enable incomplete data to effectively participate in training, thereby improving cross-center generalization ability and enabling accurate predictions for incomplete sequences in real-world clinical scenarios. To address missing sequences, we propose a cross-attention-based gated generative module (CGGM). CGGM utilizes available sequence features and learnable embedding vectors to establish cross-sequence semantic alignment through cross-attention. It then adaptively generates missing sequence features via a dynamic gating mechanism and introduces a cyclic consistency loss to constrain the semantic integrity and stability of the generated representations. The learnable embedding vectors are used to represent the global semantic distribution of the target sequence, while the gating mechanism is used to suppress uncertainty and potential noise, thereby improving the robustness of the generated features.

To fully exploit complementary information between sequences, we construct expert systems for different sequences, enabling them to simultaneously capture sequence-specific features and cross-sequence interaction features. Furthermore, we adaptively assign fusion weights based on the confidence levels of different sequence features, achieving confidence-aware fusion of features from both original and synthetic sequences. Finally, the fused global representation is fed into a multi-task classifier to simultaneously predict IDH mutation status, 1p/19q co-deletion, and pathological types. The main contributions are summarized as follows:

\begin{itemize}
\item We present GMENet, a novel network that enables accurate glioma diagnosis by effectively exploiting incomplete data, even in scenarios with missing sequences. Significantly, to the best of our knowledge, this is the pioneering study specifically dedicated to addressing incomplete imaging sequences in glioma diagnosis.

\item To address the challenge of missing sequences, we propose the CGGM to generate missing sequence features from available ones. By leveraging cross-attention and gating mechanisms, this module captures the global semantic distribution of the sequence while effectively suppressing potential noise.

\item We propose the DWEFM to explore global contextual correlations between sequence-specific and cross-sequence features. By calculating the confidence of different sequence features, this module achieves deep cross-sequence interaction and adaptive fusion.

\item  In multi-center cohorts comprising 1,241 samples, GMENet significantly increases clinically available data by 97\% (compared to complete data), improving the integrated diagnosis performance.
\end{itemize}

\section{Related Work}

\subsection{MRI-based Diagnostic Prediction of Glioma}

Recent studies have achieved notable progress in predicting key molecular markers \cite{cheng2022fully,choi2021fully,wu2022swin,van2023combined,zhang2024biologically}. Existing solutions primarily rely on either (1) multi-scale Convolutional architectures, which integrate radiomic signatures or fuse hierarchical features via backbones like ResNet and nnU-Net \cite{choi2021fully,cheng2022fully,van2023combined}, or (2) advanced modeling paradigms, which leverage Swin Transformers or multi-task learning heads to capture global dependencies and simultaneous task predictions \cite{wu2022swin,zhang2024biologically}. However, despite these architectural advancements, existing methods universally rely on complete sequences, rendering them incapable of making accurate predictions for incomplete data in real-world clinical scenarios. Given the inherent scarcity of glioma samples, this strict dependency results in a substantial waste of valuable clinical data. Moreover, most existing works have not yet aligned their predictive targets with the latest WHO CNS5 classification guidelines, thereby limiting their translational value in current precision medicine.

\begin{figure*}[!t]
	\centering
	\includegraphics[width=0.965\linewidth]{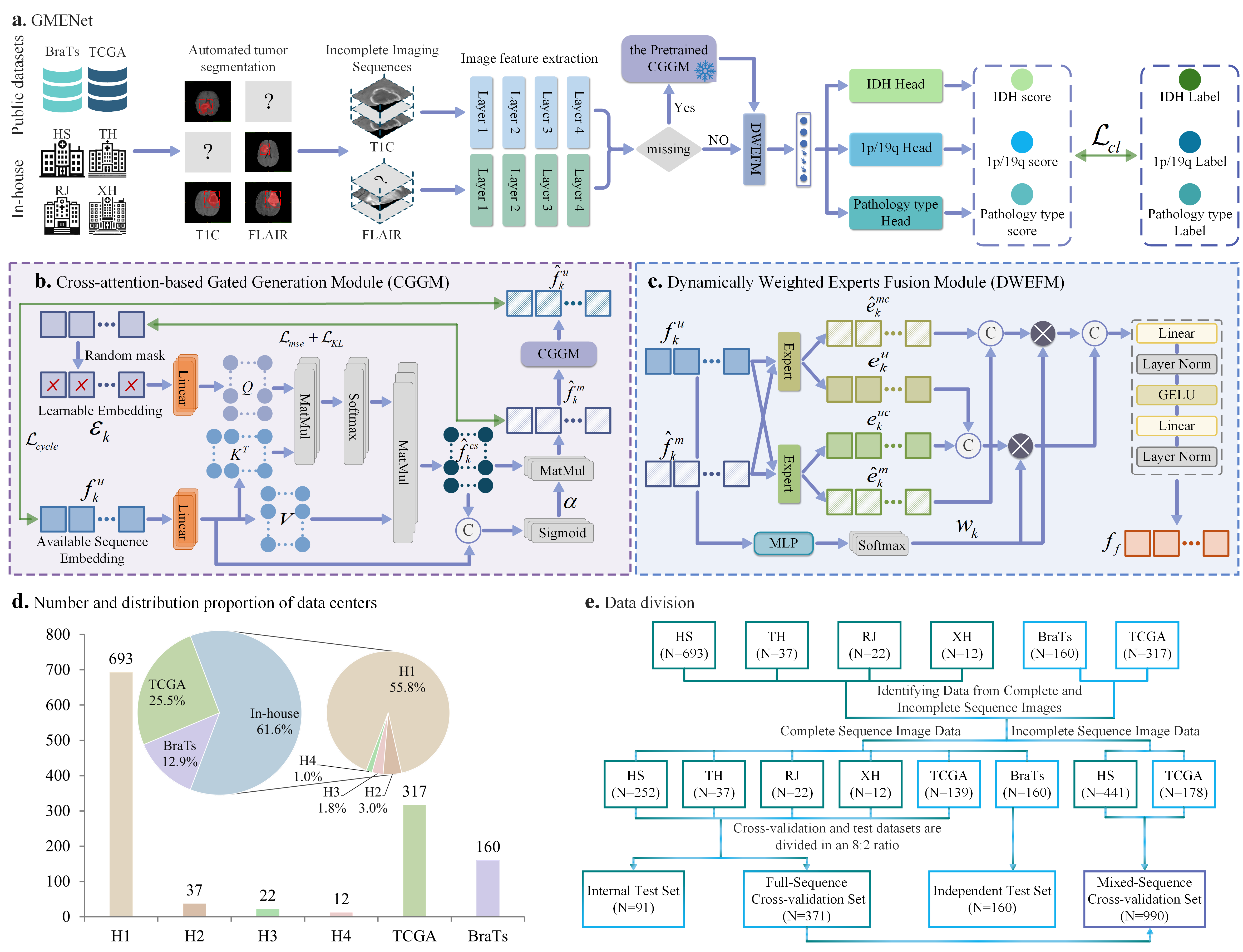}
	\caption{Schematic of the GMENet framework and data configuration. \textbf{a} The overall architecture of GMENet, utilizing the CGGM and DWEFM for simultaneous multi-task prediction on incomplete MRI data. \textbf{b} Details of the CGGM, which leverages cross-attention and gating mechanisms to synthesize missing features and suppress noise. \textbf{c} Structure of the DWEFM, employing a mixture-of-experts strategy for adaptive feature fusion. \textbf{d} Number and distribution statistics of the six data centers. \textbf{e} The dataset partitioning strategy.}
	\label{FIG:1}
\end{figure*}

\subsection{Incomplete Multimodal Learning}

Incomplete Multimodal Learning has emerged as a critical component in real-world applications \cite{xu2024leveraging,park2023cross,shi2024passion}. Prevailing strategies generally fall into two categories: (1) sequence completion via generative models (e.g., VAE \cite{shi2019variational}, GAN \cite{yoon2018gain}, Diffusion models \cite{meng2024multi}); (2) joint representation learning via cross-modal consistency \cite{lian2023gcnet,park2023cross,xu2024leveraging}. However, the former depends heavily on computationally intensive pixel-level generative architectures, while the latter often prioritizes shared representations at the expense of sequence-specific information and lacks the capability to explicitly reconstruct missing semantics, thereby limiting discriminative performance.

\section{Methodology}

\subsection{Method overview}
The architecture of GMENet is illustrated in Figure 1a. Aligning with the WHO CNS5 diagnostic criteria, GMENet is designed to simultaneously predict IDH mutation status, 1p/19q co-deletion, and pathological types. Given dual-sequence MRI images $F_k$, where $k \in \{FL, T1c\}$, we first utilize sequence-specific ResNet34 encoders to extract deep spatial features. Subsequently, a Residual MLP Stem is introduced to project the raw features from the ResNet backbone into a compact latent vector space $f_k \in R^{B \times D}$, where $B$ denotes the batch size and $D=512$ represents the feature dimension. To address the practical challenge of random missing T1c or FLAIR sequences in clinical imaging, we propose the CGGM. In this module, we introduce a learnable embedding and employ a cross-attention mechanism to synthesize missing sequence features from the existing ones. Furthermore, to ensure the semantic robustness of the synthesized features and filter out noise, we incorporate a gating mechanism aimed at generating high-quality imputed semantic features. 

When a missing imaging sequence is detected in the input data, GMENet automatically activates this pre-trained CGGM to perform real-time feature imputation. This adaptive compensation mechanism enables the network to uniformly learn deep feature representations from both complete and incomplete sequences, achieving precise diagnosis even in the presence of missing data. To effectively fuse the dual-sequence features (comprising both real and synthesized representations), we construct a DWEFM. This module adaptively assigns weights to the dual-sequence features and establishes deep cross-sequence feature interaction and refinement via a mixture-of-experts mechanism, yielding the fused feature. Finally, the fused representation is fed into a multi-task classification head to jointly predict the IDH mutation status, 1p/19q codeletion, and pathological types.

\subsection{Cross-attention-based Gated Generation Module}
To address the challenge of random sequence absence at the input stage, we propose a Cross-attention-based Gated Generation Module (CGGM), as illustrated in Figure 1b.

For a missing sequence feature, we introduce a randomly initialized learnable embedding $\mathcal{E}_k$. This vector is iteratively updated during the training process, ultimately learning the global semantic distribution prototype of the specific sequence. It serves as an anchor, guiding the model to extract features from the available sequence that align with the statistical regularities of the missing sequence, rather than merely replicating the existing features.

Drawing inspiration from prior works\cite{vaswani2017attention,xu2025mcmoe,divya2025vision}, the CGGM employs the available sequence features $f_k^u$ as Keys ($K$) and Values ($V$), and the learnable embedding $\mathcal{E}_k$ as Queries ($Q$). It utilizes a multi-head cross-attention mechanism to impute the missing sequence feature $\hat{f_k^{cs}}$. In this manner, the module learns to infer the corresponding missing sequence features from the available semantic information. The calculation process is formulated as follows:
\begin{equation}
	\hat{f_k^{cs}} = CrossAttention(\mathcal{E}_k, f_k^u, f_k^u)
\end{equation}

Subsequently, to ensure the semantic integrity of the synthesized features and mitigate the impact of noise, we incorporate a gating mechanism \cite{zhang2018gaan}. We concatenate the initially generated missing sequence feature $\hat{f_k^{cs}}$ with $f_k^u$, which is then passed through a linear layer with GELU activation to compute the gating coefficient $\alpha$. Finally, the final missing sequence feature $\hat{f_k^m}$ is obtained by modulating $\hat{f_k^{cs}}$ with $\alpha$. This mechanism ensures that the generated features are activated with high weights only when the available sequence features contain sufficient semantic cues. The computation is expressed as:
\begin{equation}
	\alpha = \sigma(Linear(Cat(\hat{f_k^{cs}}, f_k^u)))
\end{equation}
\begin{equation}
	\hat{f_k^m} = \alpha \cdot \hat{f_k^{cs}}
\end{equation}
where $\text{Cat}(\cdot)$ denotes the concatenation operation. To establish a robust feature generation space and ensure the CGGM captures authentic semantic distributions, we implement a Self-Supervised Pre-training Strategy prior to the downstream task training. Utilizing sequence-complete data, we employ a random masking strategy to simulate missing sequence feature $f_k^m$ \cite{he2022masked,tang2022self,xie2024rethinking}. To constrain the generation process, we utilize Mean Squared Error ($\mathcal{L}_{mse}$) \cite{johnson2016perceptual,zhu2024deep} and KL Divergence Loss ($\mathcal{L}_{kl}$) \cite{cui2024decoupled,liu2023fast}, targeting high-fidelity reconstruction and distributional alignment, respectively. Furthermore, to force the generator to preserve reversible semantic information, we apply a Cycle-Consistency Loss ($\mathcal{L}_{cycle}$) \cite{jang2023unsupervised,ristea2023cytran,wang2023t5}. Specifically, we utilize the generated missing sequence feature $\hat{f_k^m}$ to inversely reconstruct the original sequence feature $\hat{f_k^u}$, minimizing the discrepancy via MSE loss. The total reconstruction objective for this module is formulated as:
\begin{equation}
	\mathcal{L}_{rec} = \mathcal{L}_{mse}(\hat{f_k^m},f_k^m) + \mathcal{L}_{kl}(\hat{f_k^m},f_k^m) + \mathcal{L}_{cycle}(\hat{f_k^u},f_k^u)
\end{equation}

Upon completion of pre-training, the weights of the CGGM are frozen to preserve the learned cross-sequence reasoning capabilities during the subsequent prediction task fine-tuning.

\begin{table*}[!t]
	\setlength{\tabcolsep}{4.65pt}

	\begin{tabular}{c|crrrrrrrrrrrr}
		\toprule
		\multirow{2}{*}{ Dataset cohort}&\multirow{2}{*}{Methods} & \multicolumn{4}{c}{IDH} &\multicolumn{4}{c}{1p/19q} &  \multicolumn{4}{c}{Pathology types}  \\ 
		& &ACC	&AUC&SPE&SEN&ACC&AUC&SPE&SEN&ACC&AUC&SPE&SEN\\ \midrule
		\multirow{6}{*}{Internal Test Set}&AHI&	0.66 &	0.71 &	0.62 &	0.66 &	0.73 &	0.66 &	0.53 &	0.73 &	0.55 &	0.67 &	0.65 &	0.55 \\
		&MTTU&	0.66 &	0.69 &	0.60 &	0.66 &	0.70 &	0.66 &	0.51 &	0.70 &	0.55 &	0.67 &	0.66 &	0.55 \\
		&ST&	0.68& 	0.66 &	0.62 &	0.68 &	0.75 &	0.68 &	0.49& 	0.75 &	0.55& 	0.65 &	0.69 &	0.55 \\
		&PSNet&	0.61 &	0.60 &	0.54 &	0.61 &	0.62 &	0.59 &	0.46 &	0.62 &	0.40 &	0.55 &	0.70 &	0.40 \\
		&MDL&	0.65& 	0.71& 	0.61 &	0.65& 	0.72 &	0.66 &	0.58 &	0.72 &	0.56& 	0.70 &	0.71 &	0.56 \\
		&GMENet(FS)&	0.67 &	0.71 &	0.63 &	0.67& 	0.69 &	0.67 &	0.58& 	0.69 &	0.57 &	0.71 &	0.74& 	0.57 \\
		&GMENet&\textbf{	0.71} &	\textbf{0.76} &\textbf{	0.67} &	\textbf{0.71} &\textbf{	0.78} &	\textbf{0.69}& 	\textbf{0.61} &	\textbf{0.78} &	\textbf{0.63} &	\textbf{0.73} &\textbf{	0.76} &	\textbf{0.63} \\
		\midrule
		\multirow{6}{*}{Independent Test Set}&AHI&	0.71 &	0.77& 	0.61 &	0.71 &	0.78 &	0.54 &	0.20 &	0.78 &	0.56 &	0.76& 	0.72 &	0.56 \\
		&	MTTU&	0.77& 	0.86 &	0.69 &	0.77 &	0.80 &	0.66 &	0.27 &	0.80 &	0.59 &	0.83 &	0.79 &	0.59 \\
		&ST&	0.67 &	0.72 &	0.61& 	0.67 &	0.81 &	0.61& 	0.25 &	\textbf{0.81}& 	0.51 &	0.70 &	0.74 &	0.51 \\
		&PSNet&	0.64& 	0.69 &	0.43 &	0.64 &	0.70 &	0.58 &	0.25 &	0.70 &	0.47 &	0.63 &	0.69 &	0.47 \\
		&MDL&	0.71 &	0.79 &	0.59 &	0.71 &	0.79& 	0.51 &	0.15 &	0.79 &	0.55 &	0.70 &	0.71 &	0.55 \\
		&GMENet(FS)&	0.79 &	0.86 &	0.78& 	0.79 &	0.75 &	0.69 &	0.51 &	0.75 &	0.66 &	0.86 &	\textbf{0.83} &	0.66 \\
		&GMENet&\textbf{0.82} &	\textbf{0.90}& 	\textbf{0.77} &	\textbf{0.82} &	\textbf{0.81} &	\textbf{0.77} &	\textbf{0.56} &	\textbf{0.81} &	\textbf{0.71} &	\textbf{0.89} &	\textbf{0.83} &	\textbf{0.71} 
		
		\\ \bottomrule
	\end{tabular}
	\caption{Quantitative comparison of predictive performance across different models on the Internal Test Set and the Independent Test Set. \textbf{Bold} denotes the best result.}
	\label{tab1}
\end{table*}

\subsection{Dynamically Weighted Experts Fusion Module}

Following the imputation of missing sequence features via CGGM, we obtain a complete dual-sequence feature set $\hat{f}_k$ (comprising both real $f_k^u$ and synthesized features $\hat{f_k^m}$). To address the heterogeneity of multi-modal data and fully mine complementary information between sequences, we propose a Dynamically Weighted Experts Fusion Module (DWEFM), as illustrated in Figure 1c. This module consists of a Mixture-of-Experts (MoE) system and a dynamic routing fusion mechanism, designed to deeply refine and fuse the dual-sequence features.

\textbf{Mixture-of-Experts System:} We construct lightweight expert networks ($E_k(\cdot)$) for each sequence, which consist of Multi-layer Perceptrons with GELU activation and Layer Normalization. These independent experts are designed to specialize in capturing sequence-specific features $e_k^u$/$\hat{e_k^m}$. Furthermore, we employ these experts to process features from the alternate sequence to extract cross-sequence features $e_k^{uc}$/$\hat{e_k^{mc}}$, thereby facilitating deep cross-sequence interaction and mining implicit features invisible from a single perspective. Finally, features corresponding to the same sequence are concatenated to yield the comprehensive fused feature representation $e_k$.

\textbf{Dynamic Routing Fusion Mechanism:} To achieve adaptive weighting for different sequence features, we introduce a lightweight router $\mathcal{R}(\cdot)$. This router dynamically computes confidence weights $w_k$ based on the semantic richness of the input feature vectors $\hat{f}_k$, thereby suppressing the influence of noise features. Subsequently, we modulate $e_k$ using the confidence weights $w_k$ to generate the final refined sequence feature $e_k^w$. The calculation is formulated as follows:
\begin{equation}
	w_k = Softmax(\mathcal{R}(\hat{f}_k))
\end{equation}
\begin{equation}
	e_k^w = w_k \odot e_k
\end{equation}
where $\odot$ represents element-wise multiplication. Through this strategy, the model not only utilizes the native expert to extract standard features but also employs the cross-sequence expert to mine latent complementary information. Finally, the weighted features $e_k^w$ from both streams are concatenated and projected via a linear layer to form the fused representation $f_f$, which is enriched with cross-sequence context.

\subsection{Loss Functions}

For the joint optimization of IDH mutation status, 1p/19q co-deletion, and pathology type classification, we address the inherent long-tailed distribution of clinical data by employing the Balanced Softmax Loss ($\mathcal{L}_{bs}$) \cite{zhang2023deep,ren2020balanced,huynh2022semi}. This loss function explicitly accounts for label frequency distributions, effectively mitigating the model's bias towards majority classes. The total objective function $\mathcal{L}_{total}$ is defined as:
\begin{equation}
	\resizebox{\linewidth}{!}{$
	\mathcal{L}_{total}=\mathcal{L}_{bs}(\hat{y}_{idh},y_{idh})+\mathcal{L}_{bs}(\hat{y}_{1p/19q}, y_{1p/19q}) +\mathcal{L}_{bs}(\hat{y}_{pt}, y_{pt})$}
\end{equation}

where $\hat{y}$ and $y$ denote the predicted probabilities and ground truth labels for each task, respectively.

\section{Experiment}

\subsection{Study Population}

We retrospectively collected MRI data from 1,241 patients across six cohorts, comprising both public repositories and in-house medical institutions. The data distribution is illustrated in Figure 1d. The public datasets include the Cancer Genome Atlas (TCGA) \cite{bakas2017advancing} ($n=317$) and the Brain Tumor Segmentation (BraTS) Challenge 2020 \cite{cheng2022fully} ($n=160$), both of which provide molecular features and histopathological diagnostic information. Concurrently, data were retrospectively acquired from four in-house institutions: Ruijin Hospital, Shanghai Jiao Tong University School of Medicine (RJ, $n=22$); Xinhua Hospital, Shanghai Jiao Tong University School of Medicine (XH, $n=12$); Tongji Hospital, Tongji University (TH, $n=37$); and Huashan Hospital, Fudan University (HS, $n=693$).

\subsection{Implementation Details and Statistical Analysis}

The data division is illustrated in Figure 1e. We first identified patients with complete sequences from the multi-center cohort: BraTS ($n=160$), HS ($n=252$), TH ($n=37$), RJ ($n=22$), TCGA ($n=139$), and XH ($n=12$). The BraTS dataset ($n=160$) served as the independent test set. Data from the remaining five centers were stratified by institution and randomly split (8:2 ratio) to construct a five-fold full-sequence cross-validation (FS, $n=371$) set and an internal test set ($n=91$). Subsequently, patients with missing sequences from TCGA ($n=178$) and HS ($n=441$) were integrated into the FS set to form the mixed-sequence cross-validation (MS, $n=990$) set.

To rigorously evaluate the robustness and generalizability of the GMENet, we benchmarked it against state-of-the-art (SOTA) approaches, including AHI \cite{choi2021fully}, MTTU \cite{cheng2022fully}, ST \cite{wu2022swin}, PSNet \cite{van2023combined}, and MDL \cite{zhang2024biologically}. These methods represent representative methods emerging in recent years, which are predominantly developed based on complete datasets. In our experimental design, these comparative methods were trained exclusively on FS set. In contrast, our proposed method was trained on both MS set and FS set (denoted as GMENet(FS)). To ensure a fair comparison, all methods were evaluated on an identical test set. This comparative analysis not only underscores the clinical necessity of addressing real-world missing data but also validates the effectiveness of our proposed framework.

The proposed framework was implemented in PyTorch 1.8 and trained on a single NVIDIA A10 Tensor Core GPU using the AdamW optimizer (learning rate = $1 \times 10^{-6}$, batch size = 32). Performance was evaluated using Accuracy (ACC), Area Under the Curve (AUC), Specificity (SPE), and Sensitivity (SEN) \cite{sun2024glioma,xu2025predicting,setyawan2024beyond}. The 2D MRI images used in this study underwent a series of preprocessing steps: First, all FLAIR and T1C sequences were skull stripping and bias field correction to remove radiofrequency pulse inhomogeneity. Next, we applied the UDA-GS \cite{hu2025uda} to segment the MRI data of each patient, obtaining the segmentation masks. All results reported are the averages obtained from five-fold cross-validation set. For computational efficiency, all images were resized to $256 \times 256$, ensuring uniform aspect ratios. Data augmentation techniques, including random scaling, rotation, and normalization, were applied to enhance model robustness.
\begin{figure*}[!t]
	\centering
	\includegraphics[width=\linewidth]{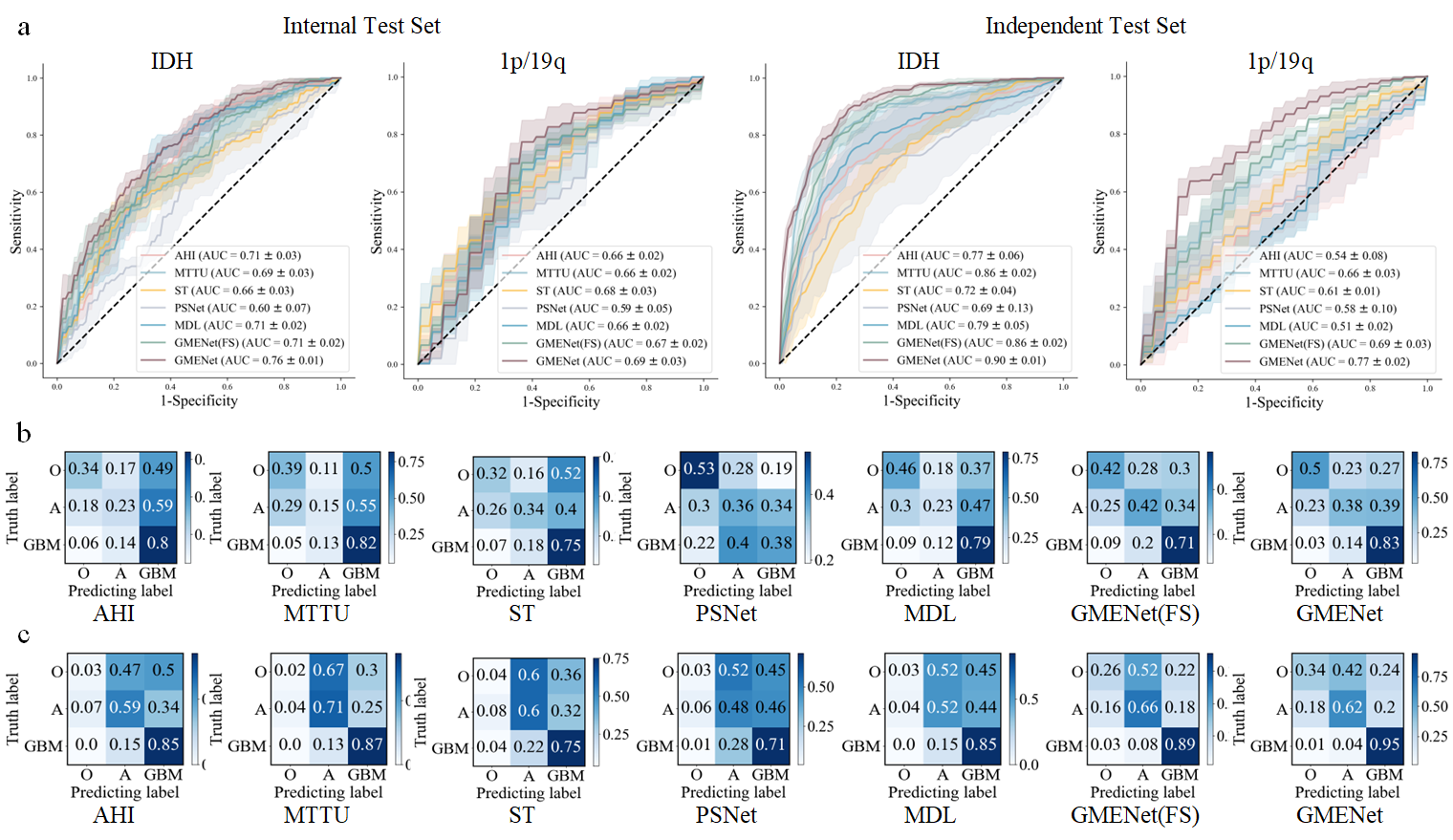}
	\caption{Performance comparison of different deep learning models on the Internal Test Set and Independent Test Set. \textbf{a} ROC curves and AUC values for molecular feature prediction on both test sets. \textbf{b} Confusion matrices for pathology type classification in the Internal Test Set. \textbf{c} Confusion matrices for pathology type classification in the Independent Test Set.
	}
	\label{FIG:2}
\end{figure*}

\subsection{Integrated Diagnostic Results for Gliomas}

The quantitative diagnostic performance of our method on both the Internal and Independent Test Sets is summarized in Table 1. The results demonstrate that even when restricted to complete sequences, GMENet(FS) exhibits highly competitive performance. In the Internal Test Set, GMENet(FS) performs on par with the competing methods in IDH prediction (AUC: 0.71) and pathology classification (ACC: 0.57), yet it demonstrates a distinct advantage in generalization capability. Notably, in the Independent Test Set, where competing methods experience significant performance degradation due to data distribution shifts, GMENet(FS) maintains remarkable robustness. It achieves a 1p/19q AUC of 0.69, significantly outperforming MTTU. Furthermore, in pathology type prediction on the Independent Test Set, GMENet(FS) achieves an accuracy of 0.66, substantially surpassing all competing approaches. This provides compelling evidence that DWEFM possesses superior generalization capabilities in feature extraction and cross-modal interaction compared to traditional feature concatenation or standard attention mechanisms.

Significantly, GMENet, trained on mixed sequences, demonstrates exceptional performance that transcends the limitations of the FS set. Despite the inclusion of incomplete sequences in the training data, GMENet consistently outperforms both GMENet(FS) and all competing methods across all three tasks. Specifically, in the Internal Test Set, GMENet elevates the AUC for IDH to 0.76 and for 1p/19q to 0.69. The ROC curves in Figure 2 (Top Row) visually corroborate this, where GMENet exhibits the largest area under the curve compared to methods such as PSNet and ST.

The confusion matrices in Figure 2 further elucidate discrepancies in class discrimination among the models. Competing methods exhibit high confusion rates between Oligodendroglioma and Astrocytoma. While GMENet(FS) achieves better class balance than competitors, GMENet further improves glioblastoma accuracy to 83\% and raises the sensitivity for Oligodendroglioma to 50\% through the mixed-sequence training strategy.

Overall, the experimental results yield two critical findings: First, our network architecture possesses superior generalization capabilities for multi-center data compared to existing SOTA methods. Second, by leveraging the CGGM and DWEFM, the model learns pathological features that are more intrinsic and robust than those acquired by models trained exclusively on complete data.
\begin{table*}[!t]
	\setlength{\tabcolsep}{3.1pt}
	\begin{tabular}{c|crrrrrrrrrrrr}
		\toprule
		\multirow{2}{*}{Dataset cohort}&\multirow{2}{*}{Methods} & \multicolumn{4}{c}{IDH} &\multicolumn{4}{c}{1p/19q} &  \multicolumn{4}{c}{Pathology types}  \\ 
		& &ACC	&AUC&SPE&SEN&ACC&AUC&SPE&SEN&ACC&AUC&SPE&SEN\\ \midrule
		\multirow{3}{*}{Internal Test Set}&GMENet(w/o CGGM)&	0.69 &	0.72 &	0.66& 	0.69 &	0.72 &	0.66 &	0.48 &	0.72 &	0.56 &	0.70& 	0.73 &	0.56 \\
		&GMENet(w/o DWEFM)&	0.66& 	0.69 &	0.61& 	0.66 &	0.73 &	\textbf{0.69} &	0.51 &	0.73 &	0.54 &	0.70 &	0.71 &	0.54 \\
		&GMENet&\textbf{0.71} &	\textbf{0.76} &	\textbf{0.67} &	\textbf{0.71} &	\textbf{0.78} &\textbf{	0.69} &	\textbf{0.61} &	\textbf{0.78} &	\textbf{0.63} &	\textbf{0.73} &	\textbf{0.76} &	\textbf{0.63} \\
		
		\midrule
		\multirow{3}{*}{Independent Test Set}&GMENet(w/o CGGM)&	0.76 &	0.86 &	0.68 &	0.76 &	0.77& 	0.73 &	0.53 &	0.77 &	0.66 &	0.84 &	0.80& 	0.66 \\
		&GMENet(w/o DWEFM)&	0.77 &	0.85& 	0.73 &	0.77 &	0.77 &	0.63 &	0.33& 	0.77 &	0.65& 	0.85 &	0.82 &	0.65 \\
		&GMENet&	\textbf{0.82}& 	\textbf{0.90} &	\textbf{0.77} &	\textbf{0.82} &\textbf{	0.81} &\textbf{	0.77}& 	\textbf{0.56} &	\textbf{0.81} &\textbf{	0.71} &	\textbf{0.89 }&\textbf{	0.83} &\textbf{	0.71} 
		\\
		\bottomrule
	\end{tabular}
	\caption{Performance comparison of models with different methodological modules on the Internal Test Set and the Independent Test Set. \textbf{Bold} denotes the best result.}
	\label{tab2}
\end{table*}
\begin{figure*}[!t]
	\centering
	\includegraphics[width=\linewidth]{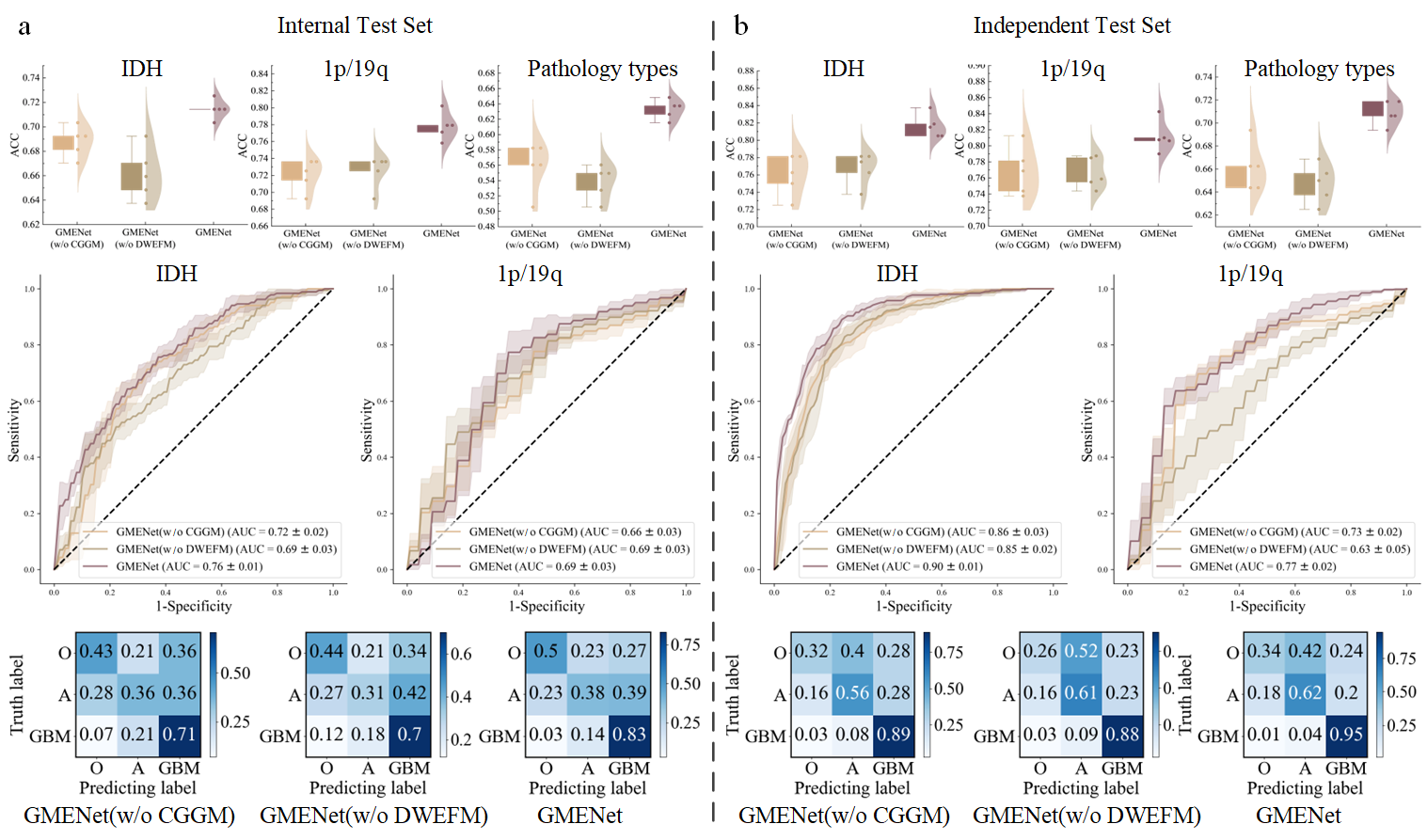}
	\caption{Performance comparison of model variants with different module configurations on the Internal Test Set (\textbf{a}) and Independent Test Set (\textbf{b}). The panels display, from top to bottom: the ACC distribution across five-fold cross-validation, ROC curves for molecular feature prediction, and confusion matrices for pathology type classification.
	}
	\label{FIG:3}
\end{figure*}

\subsection{Ablation Study}

To systematically evaluate the effectiveness of GMENet and quantify the individual contributions of its two core modules, we conducted an ablation study by removing specific modules. We evaluated the following model variants: (1) GMENet (w/o CGGM), which excludes the CGGM and substitutes missing sequences with Zero-filled inputs to assess the module's impact; (2) GMENet (w/o DWEFM), which removes the DWEFM and employs simple concatenation for feature fusion to evaluate the importance of the DWEFM; and (3) the complete GMENet. To ensure a fair comparison, all network variants were trained using identical hyperparameter settings. The quantitative results on both the Internal and Independent Test Sets are presented in Table 2 and visualized in Figure 3.

As illustrated in Table 2, the GMENet (w/o CGGM) exhibits a significant drop in predictive performance across the test sets. This decline is particularly evident in the Internal Test Set, where the AUC decreases by 4\%-10\%. Similarly, the GMENet (w/o DWEFM) also displays a severe deterioration in predictive capability, with AUC values dropping by 5\%-14\% in the Independent Test Set. As shown in the top row of Figure 3a-b, the absence of either CGGM or DWEFM not only reduces the median accuracy but also exacerbates the variance of the performance distribution, as evidenced by the wider and lower-positioned box plots. The ROC curves in the middle row of Figures 3a-b further confirm that the complete model achieves the largest AUC. Moreover, the bottom row of Figures 3a-b demonstrates that the complete GMENet significantly improves predictive performance across all tasks, achieving superior classification balance and robustness. Notably, in the Independent Test Set, GMENet boosts the sensitivity for glioblastoma to 95\% and effectively mitigates the confusion associated with oligodendroglioma.

\section{Conclusion}

In this work, we are motivated to address the pervasive challenge of incomplete imaging sequences in multi-center glioma diagnosis. Specifically, we introduced GMENet, a novel Generative Mixture of Experts Network that transforms data with missing sequences into valuable training resources, enabling precise molecular and pathological predictions in real-world clinical scenarios involving both complete and incomplete sequences. By utilizing the CGGM, GMENet dynamically synthesizes high-fidelity missing features from available sequences, effectively compensating for information scarcity via cycle-consistency constraints. Our approach further incorporates a DWEFM to adaptively integrate original and synthesized representations, maximizing the interaction and complementarity across cross-modal features. Comprehensive experiments demonstrate that GMENet can achieve reliable integrated diagnostic prediction under the clinically common condition of sequence missing data, significantly mitigating the problems caused by existing frameworks' reliance on complete sequences. Through missing sequence reconstruction and confidence-driven cross-sequence fusion, the model exhibits more stable diagnostic consistency and stronger transferability under multi-center data distribution differences.

\section*{Ethical Statement}
This study was conducted in accordance with the Declaration of Helsinki and was approved by the Institutional Ethics Committee of Huashan Hospital, Fudan University(ethics number:2015-256). The requirement for informed consent was waived by the committee due to the retrospective nature of the study.
\section*{Acknowledgements}
This work was supported by National natural science foundation of China (82372096), AI for Science Foundation of Fudan University (FudanX24AI038).

\bibliographystyle{named}
\bibliography{ijcai26}

\end{document}